\documentclass{ws-p8-50x6-00}
\usepackage{amsmath}
\usepackage{graphicx}
\def\e6{$E(6)$}
\def\10{$SO(10)$}
\def\21{$SU(2) \otimes U(1) $}

\def\422{$SU(4) \otimes SU(2) \otimes SU(2)$}
\def\321{$SU(3) \otimes SU(2) \otimes U(1)$}

\newcommand{\ed}{\end{document}}
\DeclareMathAlphabet{\mathsc}{OT1}{cmr}{m}{sc}

\def\lsim{\raise0.3ex\hbox{$\;<$\kern-0.75em\raise-1.1ex\hbox{$\sim\;$}}}
\def\gsim{\raise0.3ex\hbox{$\;>$\kern-0.75em\raise-1.1ex\hbox{$\sim\;$}}}

\newcommand{\SUSY}{\makebox[1.15cm][l]{$\line(4,1){30}$\hspace{-.95cm}{SUSY}}}
\newcommand{\tinySUSY}{\makebox[0.85cm][l]{$\line(4,1){19}$\hspace{-0.77cm}
{\tiny{SUSY}}}}

\def\simlt{\stackrel{<}{{}_\sim}}
\def\simgt{\stackrel{>}{{}_\sim}}

\newcommand{\AddrAHEP}{%
Instituto de F\'{\i}sica Te\'orica, I.F.T, \\ 
C-XVI, U.A.M., 28049 Madrid, Spain,\\}

\def\[{\left [}
\def\]{\right ]}
\def\({\left (}
\def\){\right )}

\begin{document}  

\title{A RELIEF TO THE SUPERSYMMETRIC FINE TUNING PROBLEM}

\author{J.A. Casas, J.R. Espinosa and I. Hidalgo}

\address{\AddrAHEP}

\maketitle

\abstracts{As is well known, electroweak breaking in the MSSM requires
substantial fine-tuning.  We explain why this fine tuning
problem is abnormally acute, and this allows to envisage possible
solutions to this undesirable situation. Following these ideas, we
review some  recent work which shows how in models with SUSY broken
at a  low scale (not far from the TeV) this fine-tuning can be
dramatically reduced or even absent.  }
 
\vspace{-7cm}
\rightline{IFT-UAM/CSIC-04-03}
\vspace{7cm}
\section{The abnormally acute fine tuning problem of the MSSM}
\label{sec:1}

According to general arguments, based on the size of the 
quadratically-divergent radiative corrections to the Higgs mass parameter
in the Standard Model (SM), 
the request of no  fine-tuning in the electroweak breaking 
implies that the scale of new physics should be $\Lambda\simlt$ few
TeV \cite{}. However, in the minimal supersymmetric  Standard Model (MSSM),
the absence of fine tuning requires that the masses of the new
supersymmetric particles should be  $\simlt$ few hundred
GeV. Actually, the available experimental data already imply that the
ordinary MSSM is fine tuned at least by one part in 10.  Clearly, the
fine tuning of the MSSM is abnormally acute. Let us review the
reasons for this (undesirable) situation\cite{CEH}. (For related work see
refs. \cite{BG,dCC,Poko,Anderson,Paolo,totum})

In the MSSM the Higgs 
sector
consists of two $SU(2)_L$ doublets, $H_1$, $H_2$. The 
(tree-level) scalar potential for the neutral components, $H^0_{1,2}$, 
of these doublets reads
\be
\label{VMSSM}
V^{\rm MSSM}=m_1^2|H_1^0|^2+m_2^2|H_2^0|^2- (m_3^2
H_1^0 H^0_2 + {\rm h.c.})+ \frac{1}{8}(g^2+g_Y^2) (|H^0_1|^2-|H^0_2|^2)^2,
\ee
with $m_{1,2}^2=\mu^2 + m_{H_{1,2}}^2$ and $m_3^2=B\mu$, where 
$m_{H_{i}}^2$ and $B$ are soft masses and $\mu$ is the
Higgs mass term in the superpotential, $W\supset \mu H_1\cdot  H_2$.
Minimization of $V^{\rm MSSM}$ leads to a vacuum expectation value (VEV)
$v^2\equiv 2(\langle H^0_1 \rangle^2 + \langle H^0_2\rangle^2)$
and thus to a mass for the $Z^0$ gauge boson, 
$M_Z^2=\frac{1}{4}(g^2+g_Y^2) v^2$.

The parameters of eq.(\ref{VMSSM}), in particular $m_i^2$, depend on the
initial parameters, $p_\alpha$, which for the MSSM are the 
soft masses, the $\mu-$parameter, etc. at the initial (high energy) 
scale. 
Therefore, $v^2=v^2(p_1, p_2, \cdots)$. 
The fine tuning associated to $p_\alpha$ is usually defined
by $\Delta_{p_\alpha}$ as \cite{BG}
\be
\label{BG}
{\delta v^2\over v^2} = 
\Delta_{p_\alpha}{\delta
p_\alpha\over p_\alpha}\ ,  
\ee
where $\delta v^2$ is the change induced in 
$v^2$ by a change $\delta p_\alpha$ in $p_\alpha$. 
Absence of fine tuning requires
that $\Delta_{p_\alpha}$ should not be  larger than 
${\cal O}(10)$.\footnote{Roughly speaking
$\Delta^{-1}_{p_\alpha}$ measures the probability of a cancellation
among  terms of a given size to obtain a result which is
$\Delta_{p_\alpha}$ times  smaller. For discussions see 
\cite{dCC,Poko,Anderson,Paolo}.}

Along the breaking direction in the $H_1^0, H_2^0$ space, the potential 
(\ref{VMSSM}) can be written in a 
SM-like form:
\be
\label{Vbeta}
V={1 \over 2} m^2 v^2 + {1 \over 4} \lambda v^4 \ ,
\ee
where $\lambda$ and $m^2$ are functions of the $p_\alpha$ parameters and 
$\tan\beta$ ($\equiv \langle H^0_2 \rangle/\langle H^0_1 \rangle$), in 
particular
\be
\label{cimi}
m^2 = 
c_\beta^2\, m_1^2(p_\alpha)+s_\beta^2\, m_2^2(p_\alpha)-s_{2\beta}\, 
m_3^2(p_\alpha)\ .
\ee
Minimization of (\ref{Vbeta}) leads to
\be
\label{v2}
v^2={-m^2\over \lambda} \ .
\ee
In the SM, $m^2$ is an input parameter that receives important
radiative corrections, in particular the quadratically-divergent 
ones mentioned above: $\delta m^2\propto {\Lambda^2\over 16\pi^2
v^2}(m_t^2 + \cdots)$. Hence, a tuning between the tree-level and the
one-loop contributions is required to keep $m^2$ of electroweak
size, and this sets the naturalness bound on $\Lambda$.

In the MSSM this type of corrections are absent. However, $m^2$ receives
important  logaritmic corrections $\delta m^2\propto {\tilde m^2 \over
16\pi^2}\log {M_X^2\over \tilde m^2}$, where $\tilde m$ is a  typical
soft mass and $M_X$ represents the higher scale at which the soft
breaking terms are generated. These corrections can be viewed as the
effect  of the RG running of $m^2$ from $M_X$ down to the electroweak
scale.  
Typically, the large logarithms and the numerical factors
compensate the one-loop factor, so that the corrections are quite
large, ${\cal O}(\tilde m^2)$
[actually the  tree-level values of
$m_i^2$, and thus of $m^2$, partly have a SUSY-breaking origin and
are expected to be ${\cal O}(\tilde m^2)$ as well].
This is a {\em first}
reason why the naturalness bounds on the supersymmetric masses are
more stringent than suggested by the SM  argument based on the
SM quadratically-divergent corrections
\footnote{Notice, on the other hand, that the large radiative corrections 
are usually
considered an appealing feature of the MSSM, since they trigger the
electroweak breaking in quite an elegant way, due to the negative
contribution to $m_2^2$.}.
To be concrete,
for large $\tan
\beta$ and $M_X=M_{GUT}$,
\be
\label{mMSSM}
m^2=m_1^2 c^2_\beta+m_2^2 s^2_\beta - m_3^2 s_{2\beta}
\simeq 1.01\mu^2-2.31 \tilde{m}^2\ ,
\ee
where, for simplicity, we have taken $\tilde{m}$ as the universal
value of gaugino and scalar soft masses and trilinear soft terms,
$M=m=A=\tilde{m}$. 
The presence of a sizeable RG coefficient in front of $\tilde{m}^2$ 
shows that the one-loop factor has been largely compensated.

A {\em second} (and even more important) reason for the unusual fine tuning 
of the MSSM is the following. From 
eq.(\ref{v2}),  we note that $\Delta \sim {\tt
m_i^2}/(\lambda v^2)$, where ${\tt m_i^2}$ are the (potentially large)
individual contributions to $m^2$ [see eq.(\ref{mMSSM})]. Now, 
for the MSSM $\lambda$
turns out to be quite small:
\be
\label{lambdaMSSM}
\lambda_{\rm MSSM}={1 \over 8}(g^2+g_Y^2)\cos^2 2\beta\ \simeq \  {1
\over 15}\cos^2 2\beta\ ,
\ee
which implies a fine tuning $\simgt 15$ times larger than expected
from naive dimensional considerations.

The previous $\lambda_{\rm MSSM}$ was evaluated at tree-level but radiative
corrections make $\lambda$ larger, thus reducing the fine 
tuning\cite{dCC,Poko}.  
Since $m_h^2 \sim 2\lambda v^2$, the ratio 
$\lambda_{\rm tree}/\lambda_{\rm 1-loop}$ is basically
the ratio $(m_h^2)_{\rm tree}/(m_h^2)_{\rm 1-loop}$, so for large $\tan
\beta$ the previous factor 15 is reduced by a factor
$M_Z^2/m_h^2$.  
It is important to notice that, although for a
given size of the soft terms the radiative corrections reduce the fine
tuning, the requirement of sizeable radiative corrections implies itself
large soft terms, which in turn worsens the fine tuning. More precisely,
for the MSSM $\delta_{\rm rad}\lambda\propto\log(M_{\rm SUSY}^2/m_t^2)$, 
where $M_{\rm SUSY}$ is an average of stop masses [in the universal case,
$M_{\rm SUSY}^2\simeq  3.6 \tilde m^2 + m_t^2 + 
({\rm D-terms})$]. Hence, $\lambda$ can only be radiatively enhanced 
by increasing $M_{\rm SUSY}^2$,
and thus $\tilde m$ and the individual ${\tt m_i^2}$. 
A given increase in $M_{\rm SUSY}^2$
reflects linearly in ${\tilde m}^2$ and only logarithmically in $\lambda$, so
the fine tuning $\Delta \sim {\tt m_i^2}/(\lambda v^2)$ gets usually worse. 

On the other hand, for the MSSM sizeable radiative corrections to
the Higgs mass (and thus to $\lambda$)
are in fact mandatory.
 This can be easily 
understood 
by writing the tree-level and the dominant 1-loop correction 
to the theoretical upper bound on $m_h$ in the MSSM:
\bea
\label{mhMSSM}
m_h^2\leq M_Z^2 \cos^2 2 \beta + {3 m_t^4 \over 2\pi^2 v^2}
\log{M_{\rm SUSY}^2\over m_t^2} + ...  \eea
where $m_t$ is the (running) top mass ($\simeq 167$ GeV for $M_t=174$ GeV).
Since the experimental
lower bound, $(m_h)_{\rm exp}\geq 115$ GeV, 
exceeds the tree-level contribution, the radiative
corrections must be responsible for the difference, and this translates
into a lower bound on $M_{\rm SUSY}$:
\be
\label{MSUSYMSSM}
M_{\rm SUSY}\;\simgt\;  e^{-2.1\cos^2 2\beta}
 e^{\left({m_h}/{62\ {\rm GeV}}\right)^2} m_t\;\simgt\;   3.8\ m_t\ ,
\ee
where the last figure corresponds to $m_h= 115$ GeV and large $\tan
\beta$, i.e. the most favorable case for the fine tuning. 
The last equation implies sizeable soft terms, 
$\tilde m \simgt 2 m_t$, which in turn translates into large
fine-tunings, $\Delta\simgt {\cal O}(10)$.

\begin{figure}[t]
\centerline{
\psfig{figure=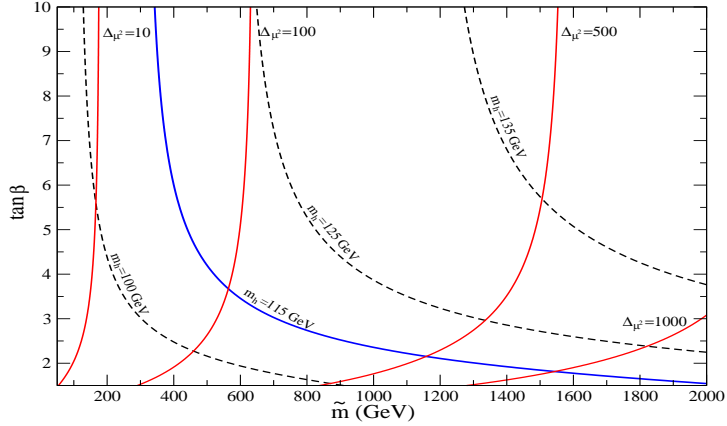,height=8cm,width=5cm,angle=-90,bbllx=5.cm,%
bblly=4.cm,bburx=20.cm,bbury=24.cm}}
\caption{\footnotesize
Fine tuning in the MSSM (measured by $\Delta_{\mu^2}$, solid lines) in the
$(\tilde{m},\tan\beta)$ plane. Dashed lines are contour lines of constant
Higgs mass.}
\label{ftMSSM}
\end{figure}

The discussion of this section  about the size of the fine-tuning in the MSSM 
is reflected in the plot of fig.\ref{ftMSSM}

\section{Possible solutions}
\label{sec:2}

As discussed above, the fine tuning of the MSSM is much more severe
than naively expected due, basically, to the smallness of the tree-level
Higgs quartic coupling, $\lambda_{\rm tree}$ and, also, to the 
large magnitude of the RG effects.
The problem is worsened by 
the fact that sizeable
radiative corrections (and thus sizeable soft terms) are needed to satisfy
the experimental bound on $m_h$. This is also due to the smallness of
$\lambda_{\rm tree}$: if it were bigger, radiative corrections would not be
necessary.  In consequence, the most efficient way of reducing the fine
tuning is to consider supersymmetric models where $\lambda_{\rm tree}$ is
larger than in the MSSM.
Then let us focus 
on $\Delta_{\mu^2}$, which can be writen 
as\footnote{$\mu^2$ is the parameter that usually 
requires the largest fine 
tuning since, due to the negative sign of its contribution in
eq.~(\ref{mMSSM}), it has to compensate the (globally
positive and large) remaining contributions.} \cite{CEH}
\bea
\label{Deltamu2}
\Delta_{\mu^2}\simeq \frac{\mu^2}{m^2}\frac{\partial
m^2}{\partial \mu^2}\simeq -\frac{\mu^2}{\lambda v^2} \simeq
-2\frac{\mu^2}{m_h^2} \ .
\eea
Strictly speaking, $m_h^2$ in 
(\ref{Deltamu2}) is the Higgs mass matrix element along the breaking 
direction, but in many cases of interest it is very close to one of the 
mass eigenvalues.
Therefore
\bea
\label{Deltamu2s}
\Delta_{\mu^2}\simeq \Delta_{\mu^2}^{\rm MSSM}
\left[\frac{m_h^{\rm MSSM}}{m_h} \right]^2 
\left[\frac{\mu}{\mu^{\rm MSSM}} \right]^2 \ .
\eea
This equation shows the two main ways in which a theory can improve
the MSSM fine tuning: increasing $m_h$ and/or decreasing $\mu$. The
first way corresponds to increasing $\lambda$. The second, for
a given $m_h$, corresponds to reducing the size of the soft
terms [from (\ref{mMSSM}) EW breaking requires the size of $\mu^2$ to be
proportional to the overall size of the soft squared-masses], which is 
only allowed if radiative contributions are not essential to raise
$m_h$. Both improvements indeed concur for larger $\lambda_{\rm 
tree}$.

The possibility of having tree-level quartic Higgs couplings larger than
in the MSSM is natural in scenarios in which the breaking of SUSY occurs
at a low-scale (not far from the TeV scale)  
\cite{hard,Brignole,Polonsky,BCEN}.\footnote{This can also happen in 
models with extra dimensions opening up not far from the electroweak scale 
\cite{Strumia}. Another way of increasing
$\lambda_{\rm tree}$ is to extend the gauge sector \cite{extG} or to
enlarge the Higgs sector \cite{extH}. The latter option has been studied
in \cite{NMSSM} (for the NMSSM) but this framework is less effective in
our opinion.} Besides, in that framework the RG effects
are largely suppressed due to the low SUSY breaking scale. As noticed above, 
this is also welcome
for the fine tuning issue. These ideas are developed in detail in the next
sections.

\section{Low-scale SUSY breaking}
\label{sec:3}

In any realistic breaking of SUSY (\SUSY), there are two scales involved: 
the \SUSY scale, say $\sqrt{F}$, which corresponds
to the VEVs of the relevant auxiliary fields in the \SUSY sector; and the
messenger scale, $M$, associated to the interactions that transmit the
breaking to the 
observable sector.
These operators give rise to soft terms (such as scalar soft masses), but
also hard terms (such as quartic scalar couplings):
\be
\label{mlambda}
m_{\rm soft}^2\sim {F^2\over M^2}\ ,\;\;\;\; \lambda_{\tinySUSY}\sim
{F^2\over M^4}\sim {m_{\rm soft}^2\over M^2}\ .
\ee
Phenomenology requires $m_{\rm soft} = {\cal O}(1\, {\rm TeV})$, but 
this does not
fix the scales $\sqrt{F}$ and $M$ separately. So, (unlike in the MSSM)
the scales $\sqrt{F}$ and $M$  could well be  of 
similar order (thus not far from the TeV scale).  This happens in the 
so-called
low-scale \SUSY scenarios\cite{hard,Brignole,Polonsky,BCEN}.  
In this framework, the hard terms of
eq.~(\ref{mlambda}), are not negligible anymore and hence the \SUSY
contributions to the Higgs quartic couplings can be easily larger than the
ordinary MSSM value (\ref{lambdaMSSM}).  As discussed in the previous 
section, this 
is
exactly the optimal situation to ameliorate the fine tuning problem.

\begin{figure}[t]
\centerline{
\psfig{figure=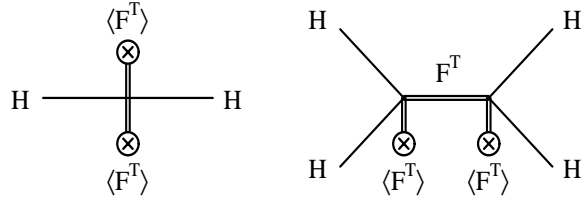,height=3cm,width=5cm,bbllx=8.cm,%
bblly=19.cm,bburx=17.cm,bbury=24.cm}}
\caption{\footnotesize  Higgs soft masses and hard quartic couplings
that arise from the K\"ahler operator discussed in sect.3.}
\label{diag}
\end{figure}

As a simple example, suppose that the K\"ahler potential contains the 
operator $K\ \supset\  -\frac{1}{M^2}|T|^2|H|^2 + \cdots$,
where $H$ denotes any
Higgs superfield and $T$ is 
the superfield responsible for \SUSY, $\langle F_T\rangle\neq 0$.
Then, 
the above nonrenormalizable
interaction produces soft terms as well as hard terms, which is 
schematically  represented in the diagrams of Fig.~\ref{diag}. 
Notice that $m_{\rm soft}^2\sim |F_T|^2/M^2$, 
$\lambda_{\tinySUSY}\sim |F_T|^2/M^4\sim 
m_{\rm soft}^2/M^2$, in agreement with (\ref{mlambda}).
More generally,
the Higgs potential has the structure 
of a generic two Higgs doublet model (2HDM), with  $T$-dependent 
coefficients. (If the $T$ field is heavy enough, it can be 
integrated out and 
one ends up with a truly 2HDM.)

The appearance of non-conventional quartic couplings has a deep impact
on the pattern of EW breaking \cite{BCEN}. In the MSSM, the existence of 
D-flat
directions, $|H_1|=|H_2|$, imposes the well-known condition, $m_1^2 +
m_2^2 - 2 |m_3^2| > 0$, in order to avoid a potential unbounded from
below along such directions. However, the boundedness of the potential can
now be simply ensured by the contribution of the extra quartic
couplings, and this opens up many new possibilities for EW breaking.  For
example, the universal case $m_1^2= m_2^2$ is now allowed, as well as
the possibility of having both $m_1^2$ and $m_2^2$ negative (with
$m_3^2$ playing a minor role).  In addition, and unlike in the MSSM, there 
is no need of radiative corrections to destabilize the origin, and EW 
breaking generically occurs already at tree-level (which is just fine
 since the effects of the RG running are
small as the cut-off scale is $M$).
Moreover, this tree-level breaking (which is welcome for the fine
tuning issue, as discussed in sect.~2) occurs naturally only in the
Higgs sector \cite{BCEN}, as desired.

Finally, the fact that quartic couplings are very different from those
of the MSSM changes dramatically the Higgs spectrum and properties.
In particular, the MSSM upper bound on the mass of the lightest
Higgs field no
longer applies, which has also an important and positive impact on
the fine tuning problem, as is clear from the discussion after 
eq.~(\ref{Deltamu2s}).

\section{A concrete model}
\label{sec:4}

In this section we evaluate numerically the fine tuning involved in the EW 
symmetry breaking in a particular model with low-scale \SUSY and compare 
it with that of the MSSM. We choose a model first introduced (as 
"example A") in 
\cite{BCEN}  and analyzed there for its own sake. We show 
now that the fine tuning problem is greatly softened in this model 
even if it was not constructed with that goal in mind.

\begin{figure}[t] 
\centerline{
\psfig{figure=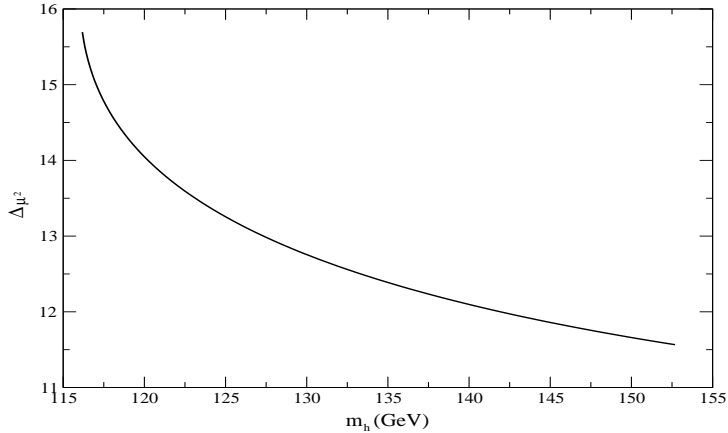,height=8cm,width=5cm,angle=-90,bbllx=5.cm,%
bblly=4.cm,bburx=20.cm,bbury=24.cm}} 
\caption{\footnotesize Fine tuning
in a low-scale SUSY breaking 
scenario as a function of the Higgs 
mass (in GeV) for $\tan\beta=10$.}
\label{ftnewvsmh2} 
\end{figure}

The superpotential is given by
\be 
W =\Lambda_S^2 T + \mu H_1\cdot H_2 + {\ell\over 2M}(H_1\cdot H_2)^2 \ ,
\ee
and the K\"ahler potential is
\bea 
K & = & |T|^2 +  |H_1|^2 +  |H_2|^2\nonumber\\ 
& - & {\alpha_t \over 4 M^2} |T|^4 + {\alpha_1 \over
M^2}|T|^2  \left(|H_1|^2+|H_2|^2\right) + {e_1 \over
2M^2}\left(|H_1|^4+|H_2|^4\right)\ .
\eea
(All parameters are real with $\alpha_t>0$.) Here $T$ is the 
singlet field responsible for the breaking of supersymmetry, $\Lambda_S$ 
is the \SUSY scale and $M$ the `messenger' scale (see previous section).
The typical soft masses are $\sim \tilde{m}\equiv \Lambda_S^2/M$. In 
particular, the mass of the scalar component of $T$ is 
${\cal{O}}(\tilde{m})$ and, after integrating this field out, the 
effective potential for $H_1$ and $H_2$ is a 2HDM  with
very particular Higgs mass terms:
\be 
\label{softm}
m_1^2=m_2^2=\mu^2-\alpha_1 \tilde{m}^2\ ,\;\;\;\; m_3^2=0\ , 
\ee
and Higgs quartic
couplings like those of the MSSM plus contributions of order $\mu/M$ and
$\tilde{m}^2/M^2$:
\bea
\label{quarticc}
\lambda_1=\lambda_2&=&{1\over 4}(g^2+g_Y^2)+2\alpha_1^2{\tilde{m}^2\over
M^2}\ ,\nonumber\\ \lambda_3&=&{1\over 4}(g^2-g_Y^2)+{2\over
M^2}(\alpha_1^2\tilde{m}^2-e_1\mu^2)\ ,\nonumber\\  \lambda_4&=&-{1\over
2}g^2-2\left(e_1+2{\alpha_1^2\over \alpha_t}\right){\mu^2\over M^2}\
,\nonumber\\  \lambda_5&=&0\ ,\nonumber\\
\lambda_6=\lambda_7&=&{\ell\mu\over M}\ .
\eea
The minimization condition for $v$ is given by eq.(\ref{v2}) with
\be
\label{dilambdaiA}
\lambda = \sum_{i=1}^7 d_i(\beta) \lambda_i(p_\alpha), \;\; \;\;\; \;
\vec{d}=({1\over 2}c^4_\beta,{1\over 2}s^4_\beta,s^2_\beta c^2_\beta,
s^2_\beta c^2_\beta,s^2_\beta c^2_\beta, 
c^2_\beta{s_{2\beta}}, s^2_\beta{s_{2\beta}})\ ,
\ee
 and there is an additional (solvable) minimization 
equation for 
$\tan\beta$
\cite{CEH}. The explicit 
expressions for $v$, $\sin2\beta$ and the spectrum of Higgs masses 
can be found in \cite{BCEN,CEH}. The corresponding expression for 
$\Delta_{\mu^2}$, as evaluated from eq.(\ref{BG}), is 
\be
\Delta_{\mu^2}=-{\mu^2\over \lambda v^2}\left[1+v^2\left(
{ls_{2\beta}\over 2\mu M}-{1\over M^2}
(e_1+{\alpha_1^2\over \alpha_t})
s_{2\beta}^2
\right)\right]\ .
\label{newfine}
\ee
To make clear the difference of behaviour with respect to the MSSM,
we plot in Fig.~\ref{ftnewvsmh2} $\Delta_{\mu^2}$ vs. $m_h$, taking 
$\mu= 330$ GeV, $\tilde{m}=550$ GeV, $e_1=-2$, $\alpha_t=1$, $l$ chosen 
to give $\tan\beta=10$, and varying
$\tilde{m}/M$ from 0.05 to 0.8. In this way we can study 
the
effect on the fine tuning of varying $\lambda$ when the low energy mass
scales ($\mu$ and $\tilde{m}$) are kept fixed. When $\tilde{m}/M$ is small
(and this implies that $\mu/M$ is also small), the unconventional
corrections to quartic couplings are not very important and the Higgs mass
tends to its MSSM value\footnote{For the model at hand this limit is not
realistic, as it implies too small (or even negative) values of
$m_A^2$, $m_H^2$ and $m_{H^\pm}^2$. However, we are interested in the 
opposite limit, of sizeable $\tilde{m}/M$.}. 
As $\tilde{m}/M$ increases, the tree level Higgs
mass (or $\lambda$) also grows and this makes $\Delta_{\mu^2}$ decrease 
with $m_h$, just the opposite of the MSSM behaviour.

Changing the parameters of this model we find many other interesting
regions, which correspond to wide
ranges of $\tan \beta$ and the Higgs masses (for more details see
ref.\cite{CEH}). Actually, the pattern of Higgs masses can be  
very different from the MSSM
and restricting the fine tuning to be less than 10 does not
impose an upper bound on the Higgs masses, in contrast with the MSSM case.
As a result, the LEP bounds do not imply a large fine tuning.
On the other hand, thanks to the size of the quartic couplings,
the Higgs mass can be as large as
several hundred GeV if desired, but this is not necessary.
 In any case, for $\Delta_{\mu^2}\leq 10$ we do find an upper bound
$\tilde{m}\simlt 500$ GeV, so that LHC would either find superpartners or
revive an (LHC) fine tuning problem for these scenarios (although the
problem would be much softer than in the MSSM).

\section{Conclusions}
\label{sec:5}

The fine tuning of the MSSM associated to the process of electroweak 
breaking is much more acute than suggested by general and intuitive 
arguments.

This is due, first, to the logaritmic corrections 
to the Higgs mass parameter, $m^2$, which
are unusually large because large logarithms 
and numerical factors compensate the one-loop suppression;
and, second (and even more important), due to the small
magnitude of the tree-level Higgs quartic coupling
$\lambda_{\rm MSSM}={1 \over 8}(g^2+g_Y^2)\cos^2 2\beta\ \simeq \  {1
\over 15}\cos^2 2\beta$. This makes the ``natural'' value for the Higgs VEV, 
$v^2 \sim m_{\rm soft}^2/\lambda$ much larger 
than $m_{\rm soft}^2$. Moreover, the smallness of $\lambda_{\rm tree}$
implies a tree-level Higgs mass smaller than the experimental lower bound.
Hence, large radiative corrections to $m_h$ (and thus
large soft terms) are required, which makes the fine tuning 
problem especially discomforting.

As a consequence, the most
efficient way of reducing the fine tuning is to consider
supersymmetric models where $\lambda_{\rm tree}$ is larger 
than in the MSSM.
This occurs naturally in scenarios in
which the breaking of SUSY occurs at a low scale (not far from the TeV
scale). As an extra bonus the radiative corrections to $m^2$
are small (EW breaking takes place at tree-level), which also helps in 
reducing the fine tuning.

We illustrate this in an explicit model, where
we achieve a dramatic
improvement of the fine tuning for any range of $\tan \beta$ and
the Higgs mass (which can be as large as several hundred GeV if 
desired, but this is not necessary).

\vspace{5pt}

\noindent
This work is supported in part by the Spanish Ministry of Science and 
Technology through a MCYT project (FPA2001-1806).
The work of I.~Hidalgo has been supported by a FPU grant from the Spanish
MECD. J.A.~Casas and J.R.~Espinosa thank the IPPP
(Durham) and the CERN TH Division respectively, for their hospitality.

\end{document}